# Design and optimisation of quantum logic circuits for a three-qubit Deutsch-Jozsa algorithm implemented with optically-controlled, solid-state quantum logic gates

**A Del Duce, S Savory and P Bayvel**
Optical Networks Group, Department of Electronic and Electrical Engineering, University College London, Torrington Place, London WC1E 7JE, UK

E-mail: a.delduce@ee.ucl.ac.uk

**Abstract.** We analyse the design and optimisation of quantum logic circuits suitable for the experimental demonstration of a three-qubit quantum computation prototype based on optically-controlled, solid-state quantum logic gates. In these gates, the interaction between two qubits carried by the electron-spin of donors is mediated by the optical excitation of a control particle placed in their proximity. First, we use a geometrical approach for analysing the entangling characteristics of these quantum gates. Then, using a genetic programming algorithm, we develop circuits for the refined Deutsch-Jozsa algorithm investigating different strategies for obtaining short total computational times. We test two separate approaches based on using different sets of entangling gates with the shortest possible gate computation time which, however, does not introduce leakage of quantum information to the control particles. The first set exploits fast approximations of controlled-phase gates as entangling gates, while the other one arbitrary entangling gates with a shorter gate computation time compared to the first set. We have identified circuits with consistently shorter total computation times when using controlled-phase gates.

## 1    Introduction

During the last years a new model of quantum computer has been developed which is based on the optically-controlled, solid-state quantum logic gates proposed by Stoneham, Fisher and Greenland in [1] and typically referred to as SFG gates[2]. In this proposal the qubits are carried by the electron-spin of donors in a solid-state substrate while two-qubit interactions are mediated by a so-called control particle placed in proximity of the qubits and triggered by the excitation and de-excitation of the control particle through optical pulses. The potential of this implementation lies in the optical control of the two-qubit interactions which allows to remove noisy electrical circuitry from the quantum register and to avoid high-precision fabrication processes for the exact placement of control electrodes.

After its first proposal presented in [1], further theoretical studies on the dynamics of SFG gates have been presented in [3], while in [2] gate parameters were identified which allow the fast implementation of entangling gates such as the C-NOT gate, for example, with minor leakage of quantum information from the qubits to the control particles. Recently, important measurements of the life-times of potential control particles in a silicon substrate have been obtained [4]. These results are a fundamental step towards the implementation of a quantum computation prototype based on SFG quantum logic gates which represents an essential test-bed for assessing the potentials of this implementation.

In this paper we contribute to this task by reporting on the design and optimisation of quantum logic circuits suitable for the experimental demonstration of a three-qubit prototype based on this technology. We start by analysing the entangling characteristics of SFG gates via the geometrical approach proposed by Zhang et al. [5]. We then address the issue of designing quantum circuits based on the SFG model which implement the Deutsch-Jozsa (DJ) algorithm [6,7]. This mathematical problem has been frequently used for testing prototypes of quantum computational systems (see, for example, [8-10]) since, it allows to demonstrate the three main features of quantum computation: parallelism, interference and entanglement [11,12]. Further, in [12], Collins et al. showed that, in a refined version of the DJ algorithm, only the implementations performed on quantum registers of at least three qubits really introduce entanglement in the system. This helped to define the minimum resources necessary to demonstrate the full power of quantum computation through this algorithm.

In our work we focus on finding the shortest possible circuits in order to reduce the chances of errors accumulating during computation. Our analysis builds on our results previously presented in [13], where a first circuit for a three-qubit refined DJ algorithm was presented. These results had been obtained adapting one of the circuits derived by Kim et al. in [10] through a generator expansion technique for an NMR quantum computer to the case of SFG quantum logic gates. However, the generator expansion technique requires the C-NOT gate to be part of the gate library used in the design process since it is exploited to reduce interactions between more than two qubits into circuits exploiting only two-qubit gates [14]. While the C-NOT gate, or the locally equivalent controlled-phase (CP) gate [5], is frequently used in literature, it is not the only entangling gate available and it is important to verify how circuit topologies may change depending on what entangling gates are used. Hence, in an effort to find the most efficient circuits, we were interested in exploiting a flexible quantum circuit design technique which would not put any particular constraints on the quantum gate library. The only constraints we considered were the ones required by universality, i.e. the gate library had to comprise two-qubit entangling gates and arbitrary single-qubit rotations [11,15], and the ones introduced by the chosen technology, i.e. the gates available for the design process had to be the ones which can be produced within the SFG model. One method that allows such flexibility to be implemented is the automated quantum circuit design algorithm based on a genetic programming approach presented by Williams and Gray [16]. Such techniques have been successfully exploited for finding quantum circuits for one- and two-qubit DJ algorithms[17,18]. However, these circuits were oracle-based, i.e. the core of the algorithm was treated as a black box. Instead, when considering to use the refined DJ algorithm for the experimental demonstration of a physical quantum computer, its complete decomposition into the one- and two-qubit gates realizable by the chosen technology is necessary.

We use Williams and Gray's approach for deriving and optimising the complete circuit sequences for a three-qubit refined DJ algorithm implemented with two-qubit SFG gates. First, we use genetic programming to find quantum circuits for a three-qubit refined DJ algorithm using ideal controlled-phase (CP) gates. With a few exceptions, we obtain very similar circuits to the ones presented by Kim et al. [10]. Then, we repeat the process exploiting SFG gates approximating fast CP gates. Finally, exploiting the ability of the SFG model to generate a large variety of different entangling gates, we further attempt to reduce the total computational time by exploiting arbitrary entangling SFG gates, meaning gates which do not have to resemble the ones typically used in literature such as the CP, C-NOT or $\sqrt{SWAP}$ gates[5,11], as long as they have the shortest possible gate operation time which, however, does not lead to loss of quantum information to the control particle. Hence, for the first time to the best of our knowledge, we present results on the efficiency of quantum circuits designed for a three-qubit refined DJ algorithm which exploit CP gates, both, ideal and approximated ones, as opposed to arbitrary entangling gates. The circuits found with our approach would be suitable for the experimental demonstration of a three-qubit quantum computation prototype based on SFG gates.

## 2   SFG quantum logic gates

In a quantum computer based on SFG quantum logic gates, the qubits are carried by the spin of an electron from a donor in a semiconductor substrate, possibly silicon [1]. Interactions between the qubits are mediated by a control particle positioned in proximity of the two qubits. In their ground states, the wavefunctions of this three-particle system are separated and no interaction between them occurs. If, however, the electron of the control particle is brought to an excited state by an optical pulse, its wavefunction spreads to the wavefunctions of the qubits, leading to an effective interaction. The interaction terminates when a second, de-exciting pulse, returns the electron of the control atom to its ground state after a time $T$. As shown in [1], of all possible $T$ values, only a discrete set defined by two integers $M$ and $N$, produces entangling gates which leave the control particle unentangled from the qubits, avoiding loss of quantum information from the proper quantum computation register to the control particles. This system can be modelled as an effective Heisenberg interaction and, assuming it to be in a magnetic field $B$ and with a strength of the exchange interaction between the control electron and qubits of $J$, the SFG entangling gate is then characterised by following parameters[1]:

$$f = \frac{B}{J} = -\frac{M^2 + N^2}{M^2 - N^2} \pm \sqrt{\left(\frac{M^2 + N^2}{M^2 - N^2}\right)^2 - 9}$$

$$JT = \frac{M\pi}{\sqrt{(f-1)^2 + 8}} = \frac{N\pi}{\sqrt{(f+1)^2 + 8}} \qquad (1)$$

$$BT = \frac{M\pi}{\sqrt{\left(1-\frac{1}{f}\right)^2 + \frac{8}{f^2}}} = \frac{N\pi}{\sqrt{\left(1+\frac{1}{f}\right)^2 + \frac{8}{f^2}}}$$

Finally, assuming the control electron to start in the spin-up state, the unitary operator describing the SFG gate is given by[3]:

$$U(M,N) = e^{i(J-B)T} \begin{bmatrix} e^{-i[(3-f)J+2B]T} & 0 & 0 & 0 \\ 0 & \frac{[(-1)^M + e^{-i(1-f)JT}]}{2} & \frac{[(-1)^M - e^{-i(1-f)JT}]}{2} & 0 \\ 0 & \frac{[(-1)^M - e^{-i(1-f)JT}]}{2} & \frac{[(-1)^M + e^{-i(1-f)JT}]}{2} & 0 \\ 0 & 0 & 0 & e^{2iBT}(-1)^N \end{bmatrix} \qquad (2)$$

From equation (2) it can be seen that a variety of different entangling gates can be produced by the SFG model. In order to better understand the entangling power of SFG gates, we have used the geometrical method proposed by Zhang et al. for the visualisation of the entangling characteristics of two-qubit gates[5]. This method is based on the observation that any two-qubit gate $U$ can be described mathematically by the expression:

$$U = k_1 e^{c_1 \sigma_{1x}\sigma_{2x} + c_2 \sigma_{1y}\sigma_{2y} + c_3 \sigma_{1z}\sigma_{2z}} k_2 \qquad (3)$$

where $k_1$ and $k_2$ are operators which only act on single qubits, and, therefore, do not influence the entangling characteristics of the gate. All the information on the entanglement power of the operator is stored in the exponential containing the three parameters $c_1$, $c_2$ and $c_3$ and the Pauli matrices $\sigma_{ix}$, $\sigma_{iy}$ and $\sigma_{iz}$, with "$i$" describing with respect to which of the two qubits the matrix is defined. The entanglement characteristics of a two-qubit gate can then be analysed by calculating $c_1$, $c_2$ and $c_3$ and by plotting them in a three-dimensional space typically called the $a^+$ Weyl chamber which contains all the entangling gates. These coefficients can be evaluated with the help of two other parameters used for comparing the entanglement of quantum gates. These are the $G_1$ and $G_2$ parameters defined by Makhlin in [26] which can be related to $c_1$, $c_2$ and $c_3$ through the expressions presented by Zhang et al. in [5]:

$$\begin{aligned} G_1 &= \cos^2(c_1)\cos^2(c_2)\cos^2(c_3) - \sin^2(c_1)\sin^2(c_2)\sin^2(c_3) \\ &\quad + \frac{i}{4}\sin(2c_1)\sin(2c_2)\sin(2c_3) \\ G_2 &= 4\cos^2(c_1)\cos^2(c_2)\cos^2(c_3) - 4\sin^2(c_1)\sin^2(c_2)\sin^2(c_3) \\ &\quad - \cos(2c_1)\cos(2c_2)\cos(2c_3) \end{aligned} \quad (4)$$

The system given in (4) can be solved for the case of SFG gates by taking the expressions of $G_1$ and $G_2$ derived specifically for the SFG model in [3]:

$$G_1(M,N) = \frac{(-1)^{(M+N)}\left[e^{-JT} + (-1)^N e^{iJT}\cos(1-f)JT\right]^2}{4} \quad (5)$$

$$G_2(M,N) = (-1)^{(M+N)}\left[\cos(2JT) + 2(-1)^N \cos(1-f)JT\right]^2$$

and noticing that expressions (4) and (5) can be respectively rearranged as:

$$\begin{aligned} 4\Re(G_1) - G_2 &= \cos(2c_1)\cos(2c_2)\cos(2c_3) \\ 4\Im(G_1) &= \sin(2c_1)\sin(2c_2)\sin(2c_3) \\ G_2 &= \cos(2c_1) + \cos(2c_2) + \cos(2c_3) \end{aligned} \quad (6)$$

and

$$\begin{aligned} 4\Re(G_1(M,N)) - G_2(M,N) &= (-1)^{(M+N)}\cos(2JT)\cos^2(JT(1-f)) \\ 4\Im(G_1(M,N)) &= -(-1)^{(M+N)}\sin(2JT)\sin^2(JT(1-f)) \\ G_2(M,N) &= (-1)^{(M+N)}\cos(2JT) + 2(-1)^M \cos^2(JT(1-f)) \end{aligned} \quad (7)$$

We evaluate the $c_1$, $c_2$ and $c_3$ parameters for SFG gates by using inspection on equations (6)-(7) obtaining:

$$\begin{aligned} c_1 &= \pi + (M+N)\frac{\pi}{2} - JT \\ c_2 &= M\frac{\pi}{2} - \frac{1}{2}JT(1-f) \\ c_3 &= M\frac{\pi}{2} - \frac{1}{2}JT(1-f) \end{aligned} \quad (8)$$

Taking expressions (8) modulo $\pi$ and through appropriate permutations, we are finally able to plot the $c_1$, $c_2$ and $c_3$ coefficients for SFG entangling gates in the $a^+$ chamber. Figure 1(a) shows the $a^+$ chamber and the $c_1$, $c_2$ and $c_3$ points plotted for $M$ and $N$ between 1 and 500. Each point represents a different entangling gate. As can also be seen from the inset in Figure 1(a), in which the points on the top have been removed in order to look inside the chamber, the points cover the whole surface of the $a^+$ chamber (with the exception of the bottom face of this tetrahedron), showing that the SFG model can implement a wide variety of different entangling gates.

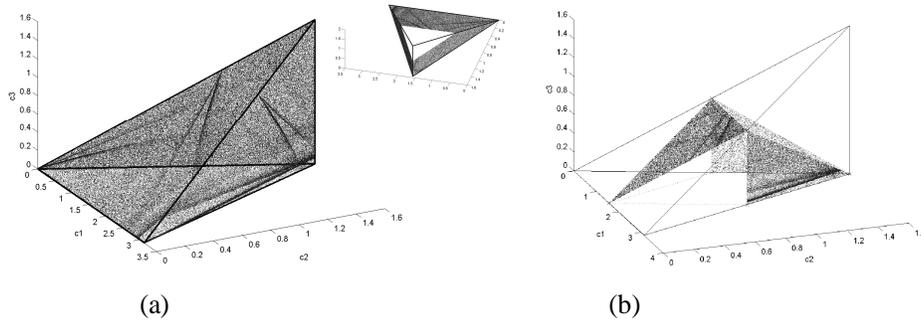

**Figure 1:(a) $c_1$, $c_2$ and $c_3$ coefficients in the $a^+$ chamber for SFG gates having *M* and *N* between 1 and 500. (b) Perfectly entangling SFG gates in the *a+* chamber.**

Further, the method proposed by Zhang et al. also allows to visualise perfect entanglers, i.e. operators able to produce a maximally entangled state from a non-entangled one. In the $a^+$ chamber, perfect entanglers are represented by operators having $c_i$ coefficients which satisfy one of the following two conditions[5]:

$$\frac{\pi}{2} \leq c_i + c_k \leq c_i + c_j + \frac{\pi}{2} \leq \pi$$
$$\frac{3\pi}{2} \leq c_i + c_k \leq c_i + c_j + \frac{\pi}{2} \leq 2\pi \quad (9)$$

where *i,j,k* are permutations of 1,2,3. The space defined by these equations and the corresponding SFG gates are shown in Figure 1(b). The ratio of the total number of entangling gates evaluated for large sets of *M* and *N* with the total number of perfect entanglers produced out of this set tends to 0.25. This is also the ratio of the total area uniformly covered by the SFG distribution with the area corresponding to perfectly entangling SFG gates, thus showing that about a quarter of all SFG gates are perfect entanglers. Now that the entangling characteristics of the SFG gate have been presented, the automated quantum circuit design tool will be described which has been developed for designing three-qubit quantum circuits based on this technology.

### 3    Genetic programming for quantum circuit design

Genetic programming for quantum circuit design has been first proposed by Williams and Gray[16]. Their model is based on following idea: suppose to have a unitary transformation $U_{comp}$ which describes a quantum computation one wishes to perform on a quantum register. Further, suppose to have a specific set of one- and two-qubit gates, described mathematically by $U_1$, $U_2$, $U_3$, $U_4…U_i$, which one is able to apply to the quantum register. This specific set of gates typically depends on the technology used to build the quantum computer under study. To perform the quantum computation corresponding to $U_{comp}$ one needs to find, i.e. design, a well-defined sequence of the implementable gates $U_i$ such that, for example:

$$U_4 \cdot U_1 \cdot U_3 \cdot U_6 \cdot U_2 \cdot U_6 = U_{comp} \quad (10)$$

Williams and Gray start their design process by arbitrarily creating an initial population of circuits. We label the number of circuits in this pool as PopL. Each circuit $U_{circ}$ comprises a random sequence of one- and two-qubit gates out of the available set:

$$U_{circ} = U_k \cdot U_m \cdot U_k \cdot U_l \cdot U_n \cdot ....... \cdot U_n \quad \{k,l,m,n\} \in \{1,2..i\} \quad (11)$$

A fitness parameter is then assigned to each circuit to quantify how well it implements the desired transformation $U_{comp}$ and the population of circuits is sorted according to their fitness. Once the initial population has been built, the algorithm continues implementing following steps:

1) *Parents* which will breed the circuits of the next generation are selected from the population. The selection procedure is random, although circuits with a higher fitness have a higher probability of being picked.
2) The next generation is created by crossover and mutation of the selected parents. In crossover, a new circuit is built by connecting two random fractions of circuits corresponding to two parents while mutation perturbs the circuit represented by a parent by, typically, inserting a random gate, deleting a random gate or perturbing an existing gate, as will be described later in more detail.
3) Once the new generation is formed, the fitness of each circuit is assessed and the population is sorted.
4) If at least one circuit in the new population has a fitness value which has reached a desired threshold or (in order to avoid excessive computation time) if a maximal number of iterations has been reached, the algorithm stops, otherwise the procedure returns to point 1).

Genetic programming involves randomly composing or mutating circuits (giving a stronger weight to ones with a high fitness level), leading to generations of better-performing circuits. Such an automated design approach is particularly helpful in the area of quantum circuit design since intuitive design approaches based on functionality of the circuits are not sufficiently developed[16].

Using genetic programming, Williams and Gray developed circuits for the teleportation problem[16]. As reviewed in [17], Spector et al. applied this method to other problems including the one- and two-qubit Deutsch-Jozsa algorithm. Other studies have analysed the impact on the search efficiency of different strategies for integrating fitness and cost functions when assessing the quality of a circuit [19] or the influence of alternative selection strategies [20]. While Williams and Gray's original work was based on finding deterministic circuits, other researchers have obtained important results limiting the search to probabilistic circuits, i.e. circuits which would yield the correct result with a probability of success p>0.5, as shown, for example, in [17,21]. In most of these studies, the quantum logic gates used for the design process were ideal and technology-independent.

Our design tool has been built specifically for the case of a three-qubit quantum register. We assume a physical distribution of the qubits such that each of the three can interact with the others as shown in Figure 2.

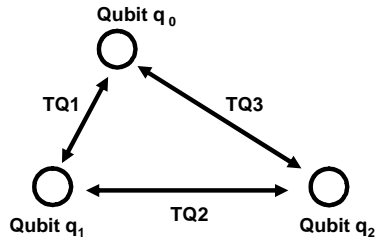

**Figure 2: Three-qubit scheme assumed for the simulations**

The set of gates we use in our design process comprises three two-qubit entangling gates TQ1, TQ2 and TQ3 representing, respectively, a two-qubit gate between qubit $q_0$ and $q_1$, $q_1$ and $q_2$, $q_0$ and $q_2$ (Figure 2). Depending on the stage of the design process, these three gates can either be three ideal and identical CP gates:

$$CP = \begin{vmatrix} 1 & 0 & 0 & 0 \\ 0 & 1 & 0 & 0 \\ 0 & 0 & 1 & 0 \\ 0 & 0 & 0 & -1 \end{vmatrix} \quad (12)$$

three different SFG gates each approximating a CP gate or three arbitrary entangling SFG gates. In terms of single-qubit gates, for each qubit we consider the $R_z(\theta)$ and $R_x(\theta)$ rotation operators and a phase shifts $Phi(\theta)$[11]:

$$R_x(\theta) = \begin{bmatrix} \cos\left(\frac{\theta}{2}\right) & -i\sin\left(\frac{\theta}{2}\right) \\ -i\sin\left(\frac{\theta}{2}\right) & \cos\left(\frac{\theta}{2}\right) \end{bmatrix}; \quad R_z(\theta) = \begin{bmatrix} e^{-i\frac{\theta}{2}} & 0 \\ 0 & e^{i\frac{\theta}{2}} \end{bmatrix}; \quad Phi(\theta) = e^{-i\frac{\theta}{2}} \begin{bmatrix} 1 & 0 \\ 0 & 1 \end{bmatrix} \quad (13)$$

These single-qubit gates together with two-qubit entangling gates form a universal set of quantum logic gates[15,11]. A circuit of $L$ quantum gates is specified through a 2x$L$ matrix in which the first row describes the type of gate and on which qubit(s) it operates while the second row stores the angles of single-qubit rotations.

When performing mutation on a circuit we consider 4 different functions. At each mutation process, a random number generator is used to select which out of the possible functions will be implemented. The four functions are: removal of a random gate, insertion of a random gate, exchange of a random gate in the circuit with a random gate from the available set and perturbation of a random gate. In the latter case, in case of two-qubit gates, we randomly change the qubits on which the selected gate is operating while in case of single-qubit operations we randomly pick a new angle $\theta$ for the rotation.

As a fitness parameter we use the average fidelity $AF$ of the transformation $U_{circ}$ produced by a given circuit with respect to the ideal transformation $U_{comp}$ to be implemented[3,22,23]:

$$AF = |Tr(U^\dagger_{comp} U_{circ})/2^{Nq}|^2 \quad (14)$$

where $Nq$ is the number of qubits in the system. $AF$ returns 1 when $U_{circ}$ implements $U_{comp}$ exactly (up to an irrelevant phase difference), while lower values are returned in case of imperfect implementation. The average fidelity has already been used to evaluate the fitness of a circuit, see, for example, [24,25], although in these cases, $AF$ was only one factor in a multi-parameter fitness function. To limit the length of the circuits in a population, we set a maximum number of two-qubit gates (TQmax) allowed for the circuits. Depending on the stage of the design process, we have used TQmax values between 3 and 20. We also exploit "cleaning" functions which condense repetitions of adjacent single-qubit gates of the same type into a single gate by adding their rotation angles.

For the selection of the parents we use stochastic universal sampling (SUS), with moderate *elitism*, i.e. the direct transfer of the best circuits to the next generation without applying crossover or mutation, to enhance the convergence of the algorithm. Our numerical tool is implemented in Matlab and the parameters key to the design process are summarised in Table 1.

**Table 1.** Main parameters of the algorithm

| Label | Parameter description |
|---|---|

| PopL | Size of the population |
| CrossProb | Fraction of the next generation bred though cross-over |
| MutProb | Fraction of the next generation bred though mutation |
| ElitProb | Fraction of the next generation bred though elitism |
| TQmax | Maximum number of two-qubit gates allowed in a circuit |

In our simulations, we used PopL=5000 while typical values for ElitProb where around 0.01, CrossProb spanned between 0.2 and 0.5 and MutProb=1-ElitProb-CrossProb.

## 4    Quantum circuits for a three-qubit refined Deutsch-Jozsa algorithm

Suppose that an oracle implements an *n*-bit function $f_n(x)$ which can be either constant, i.e. always returns either 1 or 0 for any input value, or balanced, i.e. returns 1 for exactly half of all possible input values and 0 for the remaining ones. Deutsch's problem asks how many queries to the oracle are necessary to determine whether the given function is constant or balanced. With a classical approach one needs, in the worst case, $2^{n-1}+1$ (half of all possible input values plus one) queries since, if the first $2^{n-1}$ values return the same output, say 1, one still needs one more query to define the nature of the function[6]. If the function implemented by the oracle is constant the next returned value will be again 1 whereas it will be 0 in the case of a balanced one. Instead, as reviewed in [11], Deutsch showed that using a quantum computational approach, exploiting the parallelism naturally implemented by such systems, only one query to the oracle is necessary. Further, developing a refined version of the algorithm Collins et al. showed that the circuit shown in Figure 3 correctly solves the problem [12]:

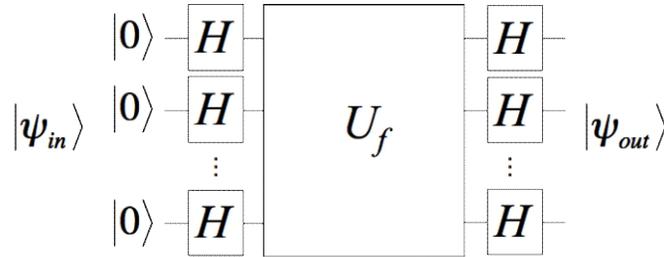

**Figure 3: Quantum circuit solving the refined Deutsch-Jozsa algorithm**

where the Hadamard gate *H* is:

$$H = \frac{1}{\sqrt{2}}\begin{vmatrix} 1 & 1 \\ 1 & -1 \end{vmatrix} \quad (15)$$

and $U_f$ is an operator which implements the function defined by the oracle. The circuit starts with an input state $|\psi_{in}\rangle$ with all *n* qubits in the $|0\rangle$-state. Then, the Hadamard gates load the qubits with an equal superposition of all possible $2^n$ values which they can represent. The operator $U_f$ is then applied to the $2^n$ input values. Finally, the last set of Hadamard gates makes the superimposed values interfere. A measurement of the output-state $|\psi_{out}\rangle$ will then find all the qubits in the $|0\rangle$-state if the implemented function was constant or return at least one qubit in the $|1\rangle$-state for the case of a balanced function.

However, to use the circuit shown in Figure 3 for the experimental demonstration of a quantum computer, one needs to find a decomposition of $U_f$ consistent with the quantum gates which the chosen technology allows to produce. Kim et al. used a generator expansion to find decompositions for all 35 non-trivial balanced functions (the constant function is implemented by replacing $U_f$ with the identity matrix) in the case of a three-qubit quantum computer exploiting NMR technology[10]. As recent studies have shown that fast and accurate CNOT gates (and therefore CP gates since they are locally equivalent to CNOTs[5]) can be produced within the SFG model[2], we start our analysis by finding circuits for the refined DJ algorithm by means of the proposed genetic programming tool using ideal CP gates as entangling gates.

*4.1 Circuits exploiting ideal CP gates*

We ran the genetic programming algorithm for all 35 balanced functions. Each function is implemented through a diagonal operator characterized by having a balanced distribution of '1s' and '-1s' on its diagonal which depends on the function's output. Since from the analysis of the results of Kim et al. in [10], we expected the angles of the $R_z$ rotations to be multiples of a fraction of $\pi$, we limited all angles in this part of the design process to be out of the ensemble $\{-\pi,-7\pi/8,...,+7\pi/8,+\pi\}$. For all functions, we obtained an exact solution, i.e. one characterised by $AF=1$, after few iterations of our genetic programming algorithm. In reporting our results, we use the same hexadecimal codification of the functions and sorting according to the number of two-qubit gates in the circuit used by Kim et al.[10]. Table 2 summarises the circuits found with our method. The subscripts of the gate labels describe to which qubit(s) the gates are applied.

**Table 2:** Quantum circuits for all 35 balanced functions obtained through genetic programming

| Function | Circuit |
|---|---|
| | 0 Two-qubit gates |
| $f_{0F}$ | $R_{2z}(\pi)$ |
| $f_{33}$ | $R_{1z}(\pi)$ |
| $f_{3C}$ | $R_{1z}(\pi)R_{2z}(\pi)$ |
| $f_{55}$ | $R_{0z}(\pi)$ |
| $f_{5A}$ | $R_{0z}(\pi)R_{2z}(\pi)$ |
| $f_{66}$ | $R_{0z}(\pi)R_{1z}(\pi)$ |
| $f_{69}$ | $R_{0z}(\pi)R_{1z}(\pi)R_{2z}(\pi)$ |
| | 1 Two-qubit gate |
| $f_{1E}$ | $R_{2z}(\pi)CP_{01}$ |
| $f_{2D}$ | $R_{0z}(-\pi)CP_{01}R_{2z}(\pi)$ |
| $f_{36}$ | $CP_{02}R_{1z}(\pi)$ |
| $f_{39}$ | $CP_{02}R_{1z}(-\pi)R_{2z}(-\pi)$ |
| $f_{4B}$ | $CP_{01}R_{0z}(\pi)R_{2z}(-\pi)$ |
| $f_{56}$ | $CP_{12}R_{0z}(-\pi)$ |
| $f_{59}$ | $R_{2z}(\pi)CP_{12}R_{0z}(-\pi)$ |
| $f_{63}$ | $CP_{02}R_{0z}(\pi)R_{1z}(\pi)$ |
| $f_{65}$ | $R_{0z}(\pi)R_{1z}(-\pi)CP_{12}$ |
| $f_{6A}$ | $R_{0z}(-\pi)R_{1z}(\pi)R_{2z}(-\pi)CP_{12}$ |
| $f_{6C}$ | $R_{1z}(-\pi)CP_{12}R_{0z}(-\pi)R_{2z}(\pi)$ |
| $f_{78}$ | $R_{1z}(-\pi)R_{2z}(\pi)CP_{12}R_{1z}(-\pi)$ |
| | 2 Two-qubit gates |
| $f_{1B}$ | $CP_{01}\,CP_{02}R_{2z}(\pi)$ |
| $f_{1D}$ | $R_{2z}(\pi)CP_{01}\,CP_{12}$ |
| $f_{27}$ | $CP_{01}\,CP_{02}\,R_{1z}(-\pi)$ |
| $f_{2E}$ | $R_{1z}(-\pi)R_{2z}(-\pi)CP_{12}\,CP_{01}$ |
| $f_{35}$ | $R_{1z}(\pi)CP_{02}\,CP_{12}$ |
| $f_{3A}$ | $CP_{12}\,R_{2z}(\pi)\,CP_{02}\,R_{1z}(\pi)$ |
| $f_{47}$ | $CP_{02}\,R_{1z}(\pi)CP_{12}$ |
| $f_{4E}$ | $R_{2z}(\pi)CP_{01}\,CP_{02}\,R_{0z}(-\pi)$ |
| $f_{53}$ | $CP_{02}\,R_{0z}(\pi)CP_{12}$ |
| $f_{5C}$ | $CP_{12}\,R_{0z}(-\pi)\,CP_{02}\,R_{2z}(-\pi)$ |
| $f_{72}$ | $R_{1z}(-\pi)CP_{01}\ R_{0z}(-\pi)CP_{02}$ |
| $f_{74}$ | $R_{1z}(-\pi)CP_{01}\ CP_{12}\ R_{0z}(\pi)$ |
| 3 Two-qubit gates | |
| $f_{17}$ | $CP_{01}\ CP_{02}\ CP_{12}$ |
| $f_{2B}$ | $CP_{12}\ CP_{02}\ CP_{01}\ R_{1z}(-\pi)R_{2z}(\pi)$ |
| $f_{4D}$ | $CP_{02}\ R_{2z}(\pi)CP_{01}CP_{12}R_{0z}(\pi)$ |
| $f_{71}$ | $CP_{12}\ R_{1z}(-\pi)CP_{01}CP_{02}R_{0z}(\pi)$ |

Comparing Table 2 and the results presented in [10], it can be seen that the circuits obtained with the two different methods require the same number of two-qubit gates. In terms of single-qubit gates, the same length of circuits has been found for all functions belonging to the group requiring 0 two-qubit gates. For the remaining functions, we found that 2D, 39, 63, 59, 65, D8, AC, CA, 27, 47, 53, 1D, 35,17 designed with our genetic programming algorithm required 1 less single-qubit gate, functions 36 and 56 two less, while function 4D required one single-qubit gate more. However, these differences might not be caused by the different methods used for the decomposition, but could also have been induced by the different entangling gates used. The gates used by Kim et al. are based on NMR technology and are locally equivalent to the CP gate, but not identical.

*4.2    Circuits exploiting SFG gates approximating CP gates*

Aiming at investigating the resources needed for the demonstration of a quantum computer based on SFG gates, we proceed analysing how the above presented circuits change when the CP gates used in the decomposition process are not ideal, but approximated via the SFG model. We continue our work presented in [13] by further developing the operator $U_{17}$ which is characterized by having the string [1 1 1 -1 1 -1 -1 -1] on its diagonal. Since at this stage the CP gates were not ideal but approximated, we allowed the angles in the $R_z$ rotations to vary continuously between $-\pi$ and $+\pi$ assuming that fixed multiples of $\pi/8$ may not be optimal anymore. Using the same gates exploited in [13], which approximated CP gates with high accuracy ($AF>0.999$), and the same static magnetic field value of 0.136meV used in [2] we used the genetic programming tool and obtained following circuit:

$$U_{17app1} = R_{z0}(-0.001)\,SFG1\,R_{z0}(0.008)\,SFG3\,R_{z1}(-0.011)\,R_{z2}(0.019)\,SFG2 \quad (16)$$

where:
-SFG1=SFG(1595,2137), acts on qubits 0 and 1
-SFG2=SFG(1584,2177), acts on qubits 1 and 2        (17)
-SFG3=SFG(815,904), acts on qubits 0 and 3

$U_{17app1}$ given in expression (16) approximates the ideal transformation $U_{17}$ with an average fidelity $AF=0.999978$. The design algorithm was stopped after about 300 rounds when no appreciable increase of the average fidelity could be observed. We can now compare $U_{17app1}$ with the circuit we previously presented in [13]. In that case, the circuit had been obtained starting from the circuit reported by Kim et al. in [10] and exploiting the local equivalence of their gates with the CP gate, leading to a final decomposition requiring 3 two-qubit gates, 7 $R_z$ rotations and 3 constant phase shifts. Instead, using the genetic programming approach we managed to save 3 $R_z$ rotations and the 3 constant phase-shifts. To assess how well a circuit implements the final state of the refined DJ algorithm we use the output-state fidelity[27]:

$$fidelity = \left|\langle \psi_{out\,ideal} | \psi_{out\,err} \rangle\right|^2 \quad (18)$$

which describes the deviation of a non-ideal output state $|\psi_{out\,err}\rangle$ (in our case, the one obtained using SFG gates) from the ideal state one wants to achieve. Expression (18) returns 1 for two equivalent states and 0 for orthogonal ones. Our new circuit obtained a fidelity value of 0.99998 compared to the 0.9998 of the previous version. Hence, the three extra $R_z$ rotations present in equation (16) compared to the shorter circuit shown in Table 2 partly compensate for the non-ideal CP gates generated by the SFG gates.

*4.3    Circuits exploiting SFG gates approximating fast CP gates*

Although the circuit given in expression (16) simulates the $U_{17}$ operator with very high precision, when considering a static magnetic field $B$ on the system of 0.136meV, one would find gate operation times $T_i$ between 80ns and 160ns for the SFG gates presented above. To reduce these gate operation times, and, therefore, the total computational time in order to protect the circuit from detrimental errors such as decoherence, for example, we have searched the entangling gates space shown in Figure 1(a) for fast SFG gates modelling CP gates whilst accepting a lower precision compared to the gates presented in (17).

Considering a static magnetic field term of 0.136meV, we found following SFG gates which approximate CP gates with $AF>0.99$ and a gate operation time $T_i<10$ns:

-SFG1=SFG(124,142), $J$=51.93GHz, $T_1$=2.63ns
-SFG2=SFG(137,156), $J$=54.37GHz, $T_2$=2.77ns     (19)
-SFG3=SFG(143,162), $J$=56.77GHz, $T_3$=2.77ns

With this set of gates, the use of the genetic programming tool helped identifying the circuit:

$$U_{17app2} = SFG2\, R_{z1}(0.038)\, SFG1\, R_{z0}(0.059)\, SFG3\, R_{z2}(0.183) \qquad (20)$$

which approximates the $U_{17}$ operator with an average fidelity of $AF$=0.9888 and the final output-state of the total refined DJ algorithm circuit with a fidelity (expression (18)) of 0.987. Again, compared to the shortest possible circuit which can be obtained with ideal CP gates, the circuit described in expression (20) uses three more single-qubit rotations. Without these, the average fidelity of the circuit implementing $U_{17}$ is 0.979, confirming that through the genetic programming tool it is possible to find single-qubit rotations which compensate for part of the non-idealities of the two-qubit gates.

*4.4    Circuits exploiting fast SFG gates*

Finally, we follow a different approach regarding the choice of gates used for designing $U_{17}$. In the results presented to this point, we have focused on circuits based on CP gates and how they could be efficiently implemented through SFG gates. However, as shown in Figure 1, there is a huge variety of entangling gates different from the CP gate which can be produced within the SFG model. Any of these entangling gates, together with single-qubit operations, forms a universal set of gates and, hence, is sufficient for designing any unitary operator. We conclude our analysis by exploring whether it is possible to design faster circuits than those presented above by exploiting arbitrary entangling gates with the shortest possible gate operation time instead of approximations of CP gates. We assume to have a three-qubit system with three random values of the exchange interaction $J$, one for each gate. Physically, this is equivalent to the situation of a semiconductor substrate hosting a random distribution of qubits and control atoms with their corresponding $J$ values given, for example, by a characterisation process[28]. Supposing to have a static magnetic field term $B$ equal to 0.136meV, we calculate the $f$ value (equation (1)) corresponding to each $J$ value and then use a continued fraction algorithm[29,2] for finding the fastest possible SFG gates which can be obtained with these parameters. In [2], this procedure had been used for finding fast C-NOT gates (which, in terms of entanglement, are equivalent to CP gates) at the expenses of having some residual entanglement between the control particle and the qubits at the end of the gate protocol. Instead, in our analysis, we focus on ideal SFG gates, in which qubits and control atoms are left unentangled at the end of a two-qubit operation, at the expenses of having slower gates compared to the ones presented in [2]. We use similar parameters as the ones used in [2] and assume for the first gate $J_1$=61.175GHz. We then arbitrarily chose $J_2$ and $J_3$ at, respectively, 5GHz and 10GHz distance from $J_1$. The continued fraction algorithm returned following gate parameters:

-SFG1=SFG(73,82), $J_1$=61.175GHz, $T_1$=1.308ns, $c_1$=1.22, $c_2$=1.22, $c_3$=0.094
-SFG2=SFG(79,88), $J_2$=66.175GHz, $T_2$=1.3055ns, $c_1$=1.311, $c_2$=1.311, $c_3$=0.0003   (21)
-SFG3=SFG(85,94), $J_3$=71.175GHz, $T_3$=1.3031ns, $c_1$=1.754, $c_2$=1.387, $c_3$=0.0071

where the $c_i$ parameters are the ones describing the location of the gates in the $a^+$ chamber. By analysing the $c_i$ parameters given in expression (21), it can be seen that the three entangling gates are different from each other and from the CP gate which has $c_i$ parameters [$\pi/2$,0,0] [5]. We ran the genetic programming tool with these gates for different maximal allowed lengths of the circuit TQmax. However, in a compromise between precision of the circuit and length, TQmax=20 was the maximum length we considered obtaining as best result the complete circuit shown in Figure 4 which also includes the Hadamard gates used at the beginning and end of the DJ algorithm.

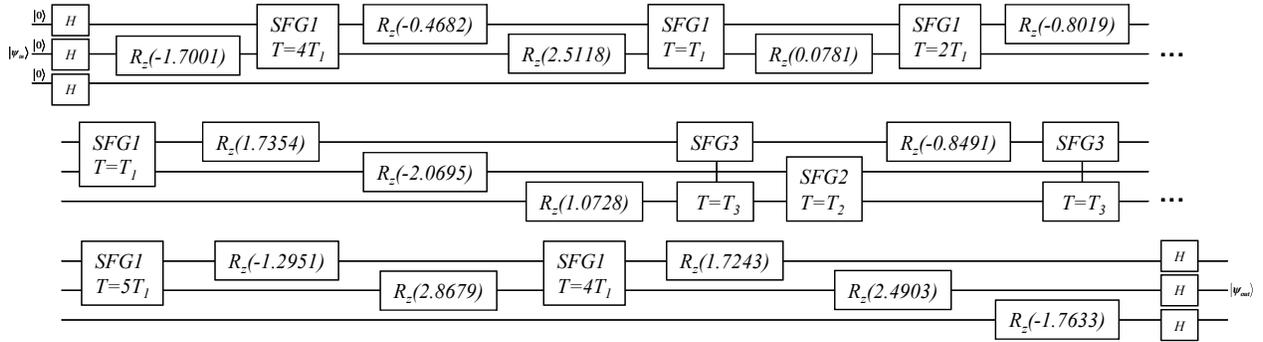

**Figure 4: Refined DJ algorithm circuit obtained with randomly entangling gates**

In Figure 4, some two-qubit gates are characterised by having a gate operation time $T=N_{rep} \cdot T_i$. This notation was used to describe sequences in which the genetic programming tool returned $N_{rep}$ repetitions of the same two-qubit gate. Remembering that SFG gates are applied through the transmission of an exciting pulse and of a de-exciting pulse after a time $T_i$, $N_{rep}$ repetitions of such a gate are equivalent to separating the exciting and de-exciting pulse by a time $N_{rep} \cdot T_i$ and experimentally corresponds to the application of a single two-qubit gate. Hence, in the circuit given in Figure 4, the implementation of $U_{17}$ requires 9 effective two-qubit gates and 14 single-qubit operations achieving an average fidelity $AF$=0.9343, while the output state of the total circuit approximates the ideal one with a fidelity (equation (18)) equal to 0.9677.

Comparing the circuit given in Figure 4, to the ones obtained using approximations of CP gates, it can be seen that, when using arbitrary entangling gates, we were only able to obtain circuits which required more than three times the number of gates required when using SFG gates modelling CP gates. In terms of computational time, the circuit given in Figure 4 requires about $20 \cdot T_i$ of time dedicated to two-qubit interactions (the sum of all the gate operation times for two-qubit gates) whereas the circuits exploiting CP gates only required about $3T_i$. Hence, despite the shorter computational time of the arbitrary entangling gates, the final circuit obtained using these gates had a longer total computational time and, moreover, achieved a lower average fidelity. We believe the reason for this to be the following. As shown in [12], all operators implementing balanced functions for a refined DJ algorithm are diagonal, with the diagonal comprising a balanced distribution of '1s' and '-1s'. The CP gate and $R_{zi}(\theta)$, are, too, diagonal operators and their multiplication returns a diagonal operator. Hence, when only using these operators, as was the case for the results presented in Table 2, the resulting circuits are very efficient. Conversely, as can be seen from expression (2), an arbitrary SFG gate has two off-diagonal elements which, once multiplied with single-qubit operators, fill off-diagonal terms of the total function operator. The design process, when using arbitrary entangling gates, has to introduce the desired sequence of '1s' and '-1s' on the diagonal and, at the same time, cancel out off-diagonal terms. Hence, although arbitrary entangling SFG gates and single-qubit operations form a universal set of gates, their structure may make the implementation of diagonal operators less efficient compared to using gates such as the CP gate. When implementing a three-qubit refined DJ algorithm, it seems therefore more efficient to choose the SFG parameters such that the corresponding entangling gates approximate CP gates, which can be done with the methods demonstrated by Kerridge et al.[2].

## 5  Conclusions

We have analysed the design of optically-controlled quantum logic circuits suitable for the experimental demonstration of a three-qubit quantum computer based on SFG gates. We have focused on the refined Deutsch-Jozsa algorithm since, despite its simplicity, it exploits the three main features of quantum computation: parallelism, interference and entanglement. In our analysis, we aimed at identifying strategies for designing the shortest possible circuits in order to protect the computation from errors introduced, for example, by decoherence. Our quantum circuit design approach was based on a genetic programming algorithm which we chose because of its flexibility with respect to the quantum gate library it uses in the decomposition process. This allowed us to compare the length of circuits obtained exploiting SFG gates approximating CP gates with circuits based on other entangling SFG gates with a shorter gate computation time. Our results showed that consistently shorter and better performing circuits can be obtained when exploiting CP gates. Further, our genetic programming tool also worked as an optimisation algorithm capable of finding single-qubit operations able to compensate part of the non-idealities which occur when using the SFG model to implement CP gates. This confirms the potential of automated quantum circuit design strategies to help overcoming inevitable imperfections which arise during the implementation of physical quantum computers.

## 6  Acknowledgments

We are grateful to Prof. A.M.Stoneham, Dr.A.Harker, Dr.P.T.Greenland and Prof. A.J.Fisher (UCL) for valuable discussions throughout the development of this work. This work was supported by RCUK and EPSRC, through the Basic Technology Programme, and the Royal Society.